\documentclass{article}%
\usepackage{graphicx}
\usepackage{amsmath}
\usepackage{amsfonts}
\usepackage{amssymb}%
\providecommand{\U}[1]{\protect\rule{.1in}{.1in}}

\begin{document}

\author{Antony Valentini\\Augustus College}

\begin{center}
{\LARGE Trans-Planckian fluctuations and the stability of quantum mechanics}

\bigskip

\bigskip

\bigskip

Antony Valentini

\textit{Department of Physics and Astronomy,}

\textit{Clemson University, Kinard Laboratory,}

\textit{Clemson, SC 29634-0978, USA.}

\bigskip

\bigskip
\end{center}

\bigskip

\bigskip

\bigskip

\bigskip

\bigskip

\bigskip

We present arguments suggesting that deviations from the Born probability rule
could be generated for trans-Planckian field modes during inflation. Such
deviations are theoretically possible in the de Broglie-Bohm pilot-wave
formulation of quantum mechanics, according to which the Born rule describes a
state of statistical equilibrium. We suggest that a stable equilibrium state
can exist only in restricted conditions: on a classical background spacetime
that is globally hyperbolic or in a mild quantum-gravity regime in which there
is an effective Schr\"{o}dinger equation with a well-defined time parameter.
These arguments suggest that quantum equilibrium will be unstable at the
Planck scale. We construct a model in which quantum nonequilibrium is
generated by a time-dependent regulator for pilot-wave dynamics, where the
regulator is introduced to eliminate phase singularities. Applying our model
to trans-Planckian modes that exit the Planck radius, we calculate the
corrected primordial power spectrum and show that it displays a power excess
(above a critical wavenumber). We briefly consider how our proposals could be
tested by measurements of the cosmic microwave background.

\bigskip

\bigskip

\bigskip

\bigskip

\bigskip

\bigskip

\bigskip

\bigskip

\bigskip

\bigskip

\bigskip

\bigskip

\bigskip

\bigskip

\bigskip

\bigskip

\bigskip

\bigskip

\section{Introduction}

According to our current understanding, the observed anisotropies in the
cosmic microwave background (CMB) were seeded by primordial quantum
fluctuations that were generated during an inflationary expansion
\cite{LL00,Muk05,W08,PU09}. Precision measurements of the CMB then allow us to
test fundamental physics at very early times and at very short distances. It
has been argued that trans-Planckian field modes -- modes with early physical
wavelengths $\lambda_{\mathrm{phys}}$ smaller than the Planck length
$l_{\mathrm{P}}$ -- are likely to contribute significantly to the inflationary
spectrum \cite{BM01,MB01}. In this case, inflationary cosmology would allow us
to probe physics at the Planck scale and beyond (for a review see ref.
\cite{BM13}).

The physical meaning of sub-Planckian lengthscales can of course be
questioned, and it may well be that an account in terms of modes with
$\lambda_{\mathrm{phys}}<l_{\mathrm{P}}$ is only an effective description.
Seen in this way, the `trans-Plankian problem' of inflationary cosmology
provides an opportunity to probe physics at the interface of quantum theory
and gravitation \cite{BM13}, an area concerning which there is as yet little
consensus and no well-established theory.

Most approaches to quantum gravity apply the standard rules of quantum
mechanics to the gravitational field
\cite{DeW67,Rov04,Thiem08,Boj11,GamPu11,Kief12}. Despite many successes,
conceptual problems remain (see for example refs. \cite{UW89,Ish91,Kuch11}).

In this paper we suggest that standard approaches to quantum gravity are based
on an implicit assumption that could turn out to be incorrect: that the
quantum-theoretical Born probability rule still holds at the Planck scale. We
suggest that this rule is relevant only in restricted conditions: on a
classical background spacetime that is globally hyperbolic, or in a mild
quantum-gravity regime with an effective time-dependent Schr\"{o}dinger
equation. In these regimes one may define a conserved quantum current in
configuration space and apply the Born rule in the usual way. But more
generally -- for example in the spacetime associated with the formation and
complete evaporation of a black hole, or in the deep quantum-gravity regime --
we suggest that there simply is no Born rule and that more general
probabilities are possible.

To make sense of this suggestion requires a formulation of quantum mechanics
in which the Born rule is not an axiom. Such a formulation is provided by the
pilot-wave theory of de Broglie and Bohm \cite{deB28,BV09,B52a,B52b,Holl93}.
In pilot-wave theory, the Born rule has a dynamical origin and is roughly
analogous to thermal equilibrium in classical physics
\cite{AV91a,AV92,AV01,VW05,EC06,TRV12,SC12,ACV14}. While the Born rule and its
empirical predictions are fully recovered in equilibrium \cite{B52a,B52b},
deviations from equilibrium and from the Born rule are theoretically possible
\cite{AV91a,AV91b,AV92,AV96,AV01,AV02,AV07,AV08,AV09,AV10,AVPwtMw,PV06}.

If such deviations existed they would generate new physics beyond the domain
of conventional quantum theory. This would include nonlocal signalling
\cite{AV91b}, which is causally consistent if one adopts an underlying
preferred foliation of spacetime \cite{AV08a}. It would also be possible to
perform `subquantum' measurements that violate the uncertainty principle and
other standard quantum constraints \cite{AV02,PV06}. On this view, quantum
physics is an effective theory of an equilibrium state and a much wider
nonequilibrium physics can exist at least in principle.

Such wider physics could have existed in the very early universe before
relaxation to equilibrium took place \cite{AV91a,AV91b,AV92,AV96}. It has been
shown that early quantum nonequilibrium can leave observable traces today, in
particular in the cosmic microwave background (CMB)
\cite{AV07,AV08,AV09,AV10,CV13,CV14,AVbook} (and perhaps in relic systems that
decoupled at very early times \cite{AV01,AV07,AV08,UnVal14}). In a cosmology
with a radiation-dominated pre-inflationary phase
\cite{VF82,L82,S82,PK07,WN08}, it is natural to expect a large-scale power
deficit in the inflationary spectrum induced by a suppression or retardation
of early relaxation at long (super-Hubble)\ wavelengths
\cite{AV07,AV08,AV10,CV13,CV14}. With appropriate cosmological parameters the
expected deficit is roughly consistent with that observed in the CMB by the
\textit{Planck} satellite \cite{PlanckXV,CV13,CV14}. While the observed
deficit may well be caused by some other more conventional effect, the fact
remains that inflationary cosmology provides us with a new and powerful
empirical window onto the Born rule in the very early universe.

In this paper we consider a different possible origin for early violations of
the Born rule. It will be suggested that quantum nonequilibrium can be
\textit{created} from an earlier equilibrium state by novel processes taking
place at the Planck scale. In addition to arguments that a stable equilibrium
state may exist only in restricted gravitational conditions (on a globally
hyperbolic spacetime or in a mild quantum-gravity regime with an effective
Schr\"{o}dinger equation), we also point out that the structure of pilot-wave
dynamics itself suggests a natural mechanism for the creation of
nonequilibrium at short lengthscales in configuration space.

Like classical general relativity, pilot-wave theory suffers from
singularities. Specifically, the de Broglie velocity field can diverge at
nodes of the wave function. To eliminate these `phase singularities', the
theory must be regularised -- an elementary point that is usually ignored
since the singularities are of measure zero (with respect to the standard
volume measure). But by including a regularisation, and allowing it to become
time-dependent, one may readily construct a simple modification of pilot-wave
dynamics in which nonequilibrium is generated from a prior equilibrium state.
This is not intended to be a fundamental theory, but only an effective or
phenomenological model of possible novel physics at the Planck scale -- in the
same spirit in which regularisation procedures in quantum field theory are not
regarded as fundamental but only as effective accounts of some unknown physics
at very short distances.

By applying our modified pilot-wave dynamics to the inflationary Bunch-Davies
vacuum, we obtain a model in which quantum nonequilibrium is created for
trans-Planckian modes as they exit the Planck radius. For a mode with
wavenumber $k$, nonequilibrium may be quantified by a function $\xi(k)$ equal
to the ratio of the nonequilibrium variance to the equilibrium (Born-rule)
variance. For a given regulator, it is possible to calculate $\xi(k)$ and so
obtain the modified primordial power spectrum, which is equal to the standard
spectrum corrected by the factor $\xi(k)$. Our model predicts a power excess
($\xi>1$) with a particular dependence on $k$. Corrections to the standard
spectrum set in above a critical wavenumber $k_{\mathrm{c}}$, set by the
comoving wavelength $\lambda_{\mathrm{c}}=2\pi/k_{\mathrm{c}}$ below which
early inflationary modes were sub-Planckian. At least in principle, best-fits
to the available CMB data could provide constraints on this kind of model,
though we do not attempt to perform such fits here.

A number of authors have proposed phenomenological or effective accounts of
trans-Planckian modifications of quantum field theory (such as modified
dispersion relations), with the aim of obtaining constraints from CMB data
\cite{BM13}. Such an approach may be taken, pending the development of a
deeper theory. In a similar spirit, we propose trans-Planckian modifications
of quantum mechanics itself. It is important to bear in mind that different
formulations of the same physics -- at the level at which the laws are
currently known -- are likely to suggest different generalisations,
modifications or extensions into a new physical domain where the laws are as
yet unknown. Thus, if one approaches the trans-Planckian domain from the
perspective of standard quantum field theory, it is natural to consider
modifications of dispersion relations, commutation relations, and other
elementary field-theoretical properties. From the perspective of the
pilot-wave formulation of quantum field theory, it natural to consider that
the Born rule may be modified -- a proposal that is not conceivable in
standard quantum field theory but which is conceptually clear in pilot-wave theory.

In Section 2 we consider our three separate arguments suggesting that the Born
rule could be unstable at the Planck scale. The first two arguments point out
that the usual derivations of a conserved probability current depend either on
the existence of a background globally-hyperbolic spacetime or on the
existence of an effective Schr\"{o}dinger equation, and that both requirements
can arguably be broken by gravitational effects. The third argument highlights
the existence of phase singularities in pilot-wave dynamics, which require
regularisation at short distances in configuration space. In Section 3 we
review the pilot-wave theory of a scalar field on expanding flat space, in
particular for the Bunch-Davies vacuum in de Sitter space. In Section 4 we
discuss our model for quantum instability, in which deviations from the Born
rule can be generated by a time-dependent regularisation of pilot-wave
dynamics. We apply this model to the Bunch-Davies vacuum and we calculate the
primordial power spectrum, which is corrected with respect to the usual result
by a factor $\xi(k)$ that generally exceeds unity. These effects can exist for
modes above a certain wavenumber $k_{\mathrm{c}}$, which underwent a Planck
radius exit some time during the inflationary phase. In Section 5 we provide a
simple estimate of $k_{\mathrm{c}}$ and discuss how the observability of the
relevant region of $k$-space depends on the values of basic cosmological
parameters. In Section 6 we summarise our conclusions and suggestions for
future work.

\section{Arguments for quantum instability at the Planck scale}

In non-gravitational physics, quantum equilibrium appears to be stable in the
sense that it is preserved in time under standard operations and interactions.
The Born rule continues to hold, for instance, in high-energy collisions (as
probed by scattering cross-sections). In pilot-wave theory this stability is a
simple consequence of the dynamics, which evolves an initial equilibrium
distribution to a final one.

A general system in pilot-wave theory has an evolving configuration $q(t)$ as
well as a wave function $\psi(q,t)$. Here $t$ is an external time parameter.
In high-energy physics, $t$ is the time associated with an underlying
preferred rest frame -- or preferred foliation of spacetime by spacelike
hypersurfaces -- and $q$ is the configuration of the fields and particles on
3-space. The Schr\"{o}dinger equation%
\begin{equation}
i\frac{\partial\psi}{\partial t}=\hat{H}\psi\label{Sch1}%
\end{equation}
(with $\hbar=1$) has an associated current $j=j\left[  \psi\right]  =j(q,t)$
in configuration space, obeying a continuity equation%
\begin{equation}
\frac{\partial\left\vert \psi\right\vert ^{2}}{\partial t}+\partial_{q}\cdot
j=0\label{Contj}%
\end{equation}
(where $\partial_{q}$ is a generalised gradient). We may then define a
pilot-wave dynamics for the system. Introducing a configuration-space velocity
field%
\begin{equation}
v(q,t)\equiv\frac{j(q,t)}{|\psi(q,t)|^{2}}\ ,\label{v}%
\end{equation}
we may write the de Broglie equation of motion%
\begin{equation}
\frac{dq}{dt}=v(q,t)\label{deB1}%
\end{equation}
for actual trajectories $q(t)$ in configuration space. Such a velocity field
$v$ exists whenever $\hat{H}$ is given by a differential operator, where the
form of $v$ (and $j$) is determined by the form of $\hat{H}$ \cite{SV09}. For
standard Hamiltonians that are quadratic in the canonical momenta, the
components $v_{a}$ of $v$ are proportional to the components of the phase
gradient:%
\begin{equation}
v_{a}\propto\partial_{q_{a}}S=\operatorname{Im}\left(  \frac{\partial_{q_{a}%
}\psi}{\psi}\right)  \ .\label{vst}%
\end{equation}
Note that the `pilot wave' $\psi$ is a complex-valued field on configuration
space that guides the motion of a single system; it has no intrinsic
connection with probability.

For an ensemble of systems with the same wave function $\psi(q,t)$ we may
consider the time evolution of an arbitrary distribution $\rho(q,t)$ of
configurations $q(t)$, where by construction $\rho(q,t)$ will obey the
continuity equation%
\begin{equation}
\frac{\partial\rho}{\partial t}+\partial_{q}\cdot\left(  \rho v\right)
=0\ .\label{cont1}%
\end{equation}
This is the same as the continuity equation (\ref{Contj}) for $\left\vert
\psi\right\vert ^{2}$. It follows that an initial distribution $\rho
(q,t_{i})=\left\vert \psi(q,t_{i})\right\vert ^{2}$ at time $t_{i}$ evolves
into a final distribution%
\begin{equation}
\rho(q,t)=\left\vert \psi(q,t)\right\vert ^{2}\label{qe1}%
\end{equation}
at time $t$.

One may also consider the time evolution of the ratio%
\begin{equation}
f\equiv\frac{\rho}{|\psi|^{2}}%
\end{equation}
along trajectories. From (\ref{cont1}) and (\ref{Contj}) it follows that%
\begin{equation}
\frac{df}{dt}=0\ ,\label{fdot}%
\end{equation}
where $d/dt=\partial/\partial t+v\cdot\partial_{q}$ is the time derivative
along a trajectory.

In the state (\ref{qe1}) of `quantum equilibrium' we obtain agreement with the
empirical predictions of quantum theory \cite{B52a,B52b}. On the other hand,
for a nonequilibrium ensemble ($\rho(q,t)\neq\left\vert \psi(q,t)\right\vert
^{2}$) the statistical predictions generally disagree with those of quantum
theory \cite{AV91a,AV91b,AV92,AV96,AV01,AV02,AV07,AV08,AV09,AV10,AVPwtMw,PV06}.

In pilot-wave dynamics the quantum equilibrium state (\ref{qe1}) is stable in
two senses: firstly, an initial equilibrium state remains in equilibrium; and
secondly, perturbations away from equilibrium tend to relax.\footnote{As shown
in ref. \cite{CV14a}, stability in this second sense does not hold for Bohm's
1952 second-order reformulation of de Broglie's original 1927 first-order
dynamics.} The relaxation process is roughly analogous to thermal relaxation
and may be quantified by the decrease of an $H$-function%
\begin{equation}
H=\int dq\ \rho\ln(\rho/\left\vert \psi\right\vert ^{2})\label{Hfn}%
\end{equation}
(on a coarse-grained level) \cite{AV91a,AV92,AV01,VW05,EC06,TRV12,SC12,ACV14}%
.\footnote{Coarse-graining is required, because of the fine-grained
conservation (\ref{fdot}) of the ratio $f$. As in the analogous classical
case, it must be assumed that the initial state has no fine-grained
micro-structure \cite{AV91a,AV92,AV01,VW05}.} Such relaxation presumably took
place in the very early universe \cite{AV91a,AV91b,AV92,AV96}.

It is the first sense of stability that concerns us here. By the above simple
reasoning, the existence of a quantum equilibrium state that is preserved by
the velocity field (\ref{v}) may be readily established for any system that
obeys a Schr\"{o}dinger equation with an associated conserved current $j$. For
example, for a bosonic scalar field $\phi$ on Minkowski spacetime we may write
quantum field theory in the functional Schr\"{o}dinger picture, with a wave
functional $\Psi\lbrack\phi,t]$, and we may assume that the velocity
$\partial\phi(\mathbf{x},t)/\partial t$ of the actual field configuration
$\phi(\mathbf{x},t)$ is given by the functional derivative $\delta
S/\delta\phi(\mathbf{x})$ where $S=\operatorname{Im}\ln\psi$ is the phase of
$\Psi$. Fermions may be described by a Dirac-sea picture, with particle
trajectories determined by a many-body Dirac wave function
\cite{BH93,C03,CS07}. These constructions require a preferred time parameter
$t$ with respect to which nonlocal effects (arising from the equation of
motion (\ref{deB1}) for entangled quantum states) occur instantaneously. For
ensembles of fields or particles in quantum equilibrium, we recover standard
quantum field theory (and hence an effective Lorentz invariance)
\cite{BHK87,AVbook}.

In the absence of gravitation, then, the existence of a quantum equilibrium
state is a trivial consequence of the structure of pilot-wave dynamics. In the
presence of gravitation, however, the situation is not so clear. We shall now
present arguments suggesting that quantum equilibrium may in fact be
gravitationally unstable.

\subsection{Globally-hyperbolic spacetime and the existence of a quantum
equilibrium state}

The existence of a quantum equilibrium state may be readily established on a
classical curved spacetime background that is globally hyperbolic \cite{AV04}.

Such a spacetime may always be foliated (in general nonuniquely) by spacelike
hypersurfaces $\Sigma(t)$ that are labelled by a global time function $t$. The
spacetime line element $d\tau^{2}=\,^{(4)}g_{\mu\nu}dx^{\mu}dx^{\nu}$ with
4-metric $^{(4)}g_{\mu\nu}$ may then be written in the standard 3+1 form%
\begin{equation}
d\tau^{2}=(N^{2}-N_{i}N^{i})dt^{2}-2N_{i}dx^{i}dt-g_{ij}dx^{i}dx^{j}\ ,
\label{ADM}%
\end{equation}
where $N$ is the lapse function, $N^{i}$ is the shift vector and $g_{ij}$ is
the 3-metric on $\Sigma(t)$. We may set $N^{i}=0$ (for as long as the lines
$x^{i}=\mathrm{const}.$, chosen to be normal to the slices $\Sigma$, do not
encounter singularities).

For example, for a massless and minimally-coupled real scalar field $\phi$
with Lagrangian density%
\begin{equation}
\mathcal{L}=\frac{1}{2}\sqrt{-\,^{(4)}g}\,^{(4)}g^{\mu\nu}\partial_{\mu}%
\phi\partial_{\nu}\phi
\end{equation}
we have a canonical momentum density $\pi=\partial\mathcal{L}/\partial
\dot{\phi}=(\sqrt{g}/N)\dot{\phi}$ (where $^{(4)}g=\det g_{\mu\nu}$ and
$g=\det g_{ij}$) and a classical Hamiltonian%
\begin{equation}
H=\int d^{3}x\;\frac{1}{2}N\sqrt{g}\left(  \frac{1}{g}\pi^{2}+g^{ij}%
\partial_{i}\phi\partial_{j}\phi\right)  \ .
\end{equation}
The wave functional $\Psi\lbrack\phi,t]$ then satisfies the Schr\"{o}dinger
equation\footnote{As usual in this context, we implicitly assume some form of
regularisation -- such as an analytical continuation of the number of space
dimensions away from 3 (see, for example, ref. \cite{Guven1989}).}%
\begin{equation}
i\frac{\partial\Psi}{\partial t}=\int d^{3}x\;\frac{1}{2}N\sqrt{g}\left(
-\frac{1}{g}\frac{\delta^{2}}{\delta\phi^{2}}+g^{ij}\partial_{i}\phi
\partial_{j}\phi\right)  \Psi\ . \label{Sch2}%
\end{equation}
This implies a continuity equation%
\begin{equation}
\frac{\partial\left\vert \Psi\right\vert ^{2}}{\partial t}+\int d^{3}%
x\;\frac{\delta}{\delta\phi}\left(  \left\vert \Psi\right\vert ^{2}\frac
{N}{\sqrt{g}}\frac{\delta S}{\delta\phi}\right)  =0 \label{cont2}%
\end{equation}
with a current $j=\left\vert \Psi\right\vert ^{2}(N/\sqrt{g})\delta
S/\delta\phi$ and a de Broglie velocity field%
\begin{equation}
\frac{\partial\phi}{\partial t}=\frac{N}{\sqrt{g}}\frac{\delta S}{\delta\phi
}\ , \label{deB2}%
\end{equation}
where $\Psi=\left\vert \Psi\right\vert e^{iS}$ \cite{AV04,AVbook}.

The field velocity (\ref{deB2}) at a point $x^{i}$ on $\Sigma(t)$ will depend
instantaneously (with respect to $t$) on field values at distant points
$(x^{\prime})^{i}\neq x^{i}$ -- if $\Psi$ is entangled with respect to the
fields at those points. For a nonequilibrium ensemble, a change in the local
Hamiltonian at $(x^{\prime})^{i}$ will in general instantaneously affect the
time evolution of the marginal distribution at $x^{i}$, yielding nonlocal
signals from $(x^{\prime})^{i}$ to $x^{i}$ (whereas in equilibrium such
signals will vanish) \cite{AV91b}. To ensure physical consistency, we assume
that the theory has been constructed using a preferred foliation associated
with a specific lapse function $N(x^{i},t)$ \cite{AV08a}.

By construction, an arbitrary distribution $P[\phi,t]$ will satisfy the same
continuity equation:%
\begin{equation}
\frac{\partial P}{\partial t}+\int d^{3}x\;\frac{\delta}{\delta\phi}\left(
P\frac{N}{\sqrt{g}}\frac{\delta S}{\delta\phi}\right)  =0\ . \label{cont2'}%
\end{equation}
It then follows as usual that $P[\phi,t]=\left\vert \Psi\lbrack\phi
,t]\right\vert ^{2}$ is an equilibrium state: if it holds at some initial time
it will hold at all times. Thus there is a quantum equilibrium state even in
the presence of gravitation -- at least for a classical curved spacetime
background that is globally hyperbolic.

It is however difficult to see how a comparable construction could be given
for a background spacetime that is not globally hyperbolic -- such as the
spacetime generated by the formation and (complete) evaporation of a black
hole \cite{Hawk76}. Even standard quantum field theory on curved spacetime
relies on the assumption that the spacetime is globally hyperbolic. The usual
quantisation procedure imposes canonical commutation relations on a Cauchy
surface, so that the wave equation has a well-posed initial value formulation
(see for example ref. \cite{Wald94}). In effect the standard theory depends on
the quantisation of a well-posed Hamiltonian dynamics for classical fields,
and is therefore strictly speaking applicable only to globally-hyperbolic
spacetimes. An algebraic approach to quantum field theory on
non-globally-hyperbolic spacetimes has been developed and applied to simple,
flat (two-dimensional) examples \cite{Yurt94}. In this construction, the
algebraically-defined quantum state must be specified on the entire spacetime
with boundary conditions at naked singularities. It is unclear if this
approach could be given a de Broglie-Bohm formulation.

Existing pilot-wave theories require a preferred hypersurface along which
nonlocality acts \cite{AV08a}. Even in flat spacetime, attempts to write down
a fundamentally Lorentz-invariant pilot-wave theory run into problems
associated with nonlocality: both the dynamics and the quantum equilibrium
distribution must be defined on a preferred spacelike hypersurface
\cite{Hardy92,BerndlGold94,Berndletal96,Myrv02}.\footnote{Ref.
\cite{Duerretal14} considers a new velocity law (replacing (\ref{deB1}))
involving a non-integrable time-like vector field $n^{\mu}$, with the aim of
formulating a fundamentally Lorentz-invariant pilot-wave theory. However,
because the model has no well-defined foliation (or global time function) the
equilibrium state is ill-defined -- except when $n^{\mu}$ happens to be
integrable and determines a preferred foliation.} In the absence of a Cauchy
hypersurface, we may expect a fundamental difficulty in defining a quantum
equilibrium state for a nonlocal hidden-variables theory.

The de Broglie-Bohm construction depends on the existence of a local quantum
current in configuration space, and there seems to be no reason why such a
current would exist for a non-globally hyperbolic spacetime. Pending a
demonstration to the contrary, one may consider the possibility that such a
current does not exist. It has in fact been suggested that there is no
well-defined state of quantum equilibrium for a non-globally hyperbolic
spacetime and that the formation and complete evaporation of a black hole
could generate quantum nonequilibrium from a prior equilibrium state
\cite{AV04,AV07}. If such effects existed, outgoing Hawking radiation would be
in a state of quantum nonequilibrium and could therefore carry more
information than ordinary radiation -- potentially offering a new approach to
the (controversial) question of information loss in black holes
\cite{AV04,AV07}.

\subsection{Possible \textbf{non-existence of an equilibrium state} in quantum
gravity}

The existence of a quantum equilibrium state is difficult to establish in
canonical quantum gravity because of the absence of a general time-dependent
Schr\"{o}dinger equation with an associated probability current. Here we
suggest that it may in fact be a mistake to assume that there exists a state
of quantum equilibrium at the Planck scale.

Canonical quantum gravity begins with the Einstein-Hilbert action%
\[
I=\int d^{4}x\ (-^{(4)}g)^{1/2}\ ^{(4)}R
\]
(units $G=1/16\pi$). Employing the standard 3+1 splitting (\ref{ADM}), the
arbitrariness of the lapse function $N$ implies the Wheeler-DeWitt equation
\cite{DeW67}%
\begin{equation}
\left(  -G_{ijkl}\frac{\delta^{2}}{\delta g_{ij}\delta g_{kl}}-g^{1/2}%
R\right)  \Psi=0\ ,\label{WD}%
\end{equation}
where $G_{ijkl}={\frac{1}{2}}g^{-1/2}(g_{ik}g_{jl}+g_{il}g_{jk}-g_{ij}g_{kl})$
is the superspace metric, $R$ is the 3-scalar curvature, and we employ the
metric representation with a wave functional $\Psi=\Psi\lbrack g_{ij}]$.
(Further constraints expressing spatial diffeomorphism invariance on $\Sigma$
read $\left(  \delta\Psi/\delta g_{ij}\right)  _{;\,j}=0$, where $;\,j$ is the
3-covariant derivative.)

Many applications of (\ref{WD}) to quantum cosmology make use of a
semiclassical WKB approach, where one writes $\Psi=\left\vert \Psi\right\vert
e^{iS}$ -- with $\left\vert \Psi\right\vert $ varying slowly with $g_{ij}$ --
and assumes approximately classical trajectories for $g_{ij}$ given by
$p^{ij}=\delta S/\delta g_{ij}$, where $p^{ij}$ is the momentum density
canonically-conjugate to $g_{ij}$. Using the canonical relation between
$p^{ij}$ and $\dot{g}_{ij}$, we then have an equation of motion%
\begin{equation}
\frac{\partial g_{ij}}{\partial t}=2NG_{ijkl}\frac{\delta S}{\delta g_{kl}%
}+N_{i\,;\,j}+N_{j\,;\,i}\ .\label{deBWD}%
\end{equation}

The trajectories defined by (\ref{deBWD}) are in effect de Broglie-Bohm
trajectories, for the special case of a WKB wave functional. A number of
authors have proposed a pilot-wave formulation of quantum gravity based on
(\ref{WD}) and (\ref{deBWD}) -- with (\ref{deBWD}) assumed to be valid for any
solution $\Psi\lbrack g_{ij}]$ of (\ref{WD}) \cite{Vink92,Holl93,Shtanov96}.
The resulting theory has been extensively applied to quantum cosmology (for a
review see ref. \cite{PNF13}).

At the level of individual systems, an important question concerns the
consistency of the pilot-wave dynamics defined by (\ref{WD}) and
(\ref{deBWD}). The lapse and shift functions $N$ and $N^{i}$ are arbitrary, so
any change in these should not affect the resulting 4-geometry traced out by
the evolution of the 3-geometry. Otherwise the initial-value problem would be
ill-posed. Shtanov \cite{Shtanov96} gave an example for which it appeared that
the predicted 4-geometry would depend on the arbitrary choice of lapse
function $N$. This was interpreted as a breakdown of foliation invariance. One
might then just as well abandon foliation invariance from the outset, and
adopt a time-dependent Schr\"{o}dinger equation with a specific choice of
lapse $N$ and a preferred time parameter $t$ (as suggested in a de
Broglie-Bohm context in refs. \cite{AV92,AV96,AVbook,RosVal14}).

But the work of Pinto-Neto and Santini \cite{PNS02} appears to demonstrate
that the above pilot-wave dynamics with the Wheeler-DeWitt equation is in fact
well-posed. Pinto-Neto and Santini rewrite the dynamics of the de Broglie-Bohm
trajectories in a classical Hamiltonian form. This is done by adding a term of
the form $Nq$ to the classical Hamiltonian density where%
\begin{equation}
q=-\frac{1}{|\Psi|}G_{ijkl}\frac{\delta^{2}|\Psi|}{\delta g_{ij}\delta g_{kl}%
}\ .
\end{equation}
Given the guidance equation $p^{ij}=\delta S/\delta g_{ij}$ at an initial
time, Hamilton's equations then generate the same de Broglie-Bohm trajectories
as would be generated by the guidance equation applied at all times. By
applying well-known theorems \cite{Dirac64,Teit73,HKT76} it then possible to
show that, given (consistent) initial conditions on a spacelike slice, the
resulting 4-geometry is independent of the choice of $N$ and $N^{i}$. For
$q\neq0$, while the algebra of constraints is closed (when evaluated on the
trajectories) it differs from the classical Dirac-Teitelboim algebra.
Pinto-Neto and Santini conclude that, while the time evolution is consistent,
in general it will form a spacetime with a non-Lorentzian structure (a
degenerate 4-geometry) -- unless $q$ happens to vanish, in which case one
recovers a classical evolution and a locally-Lorentzian spacetime. The
breaking of the Dirac-Teitelboim algebra for $q\neq0$ is interpreted as a
breaking of local Lorentz invariance at the level of individual trajectories,
caused by the nonlocality associated with $q\neq0$.

However, if the pilot-wave dynamics of the Wheeler-DeWitt equation is indeed
well-posed as a dynamical theory of a 3-geometry evolving in time, the
question remains of how to connect the dynamics of single systems to the
theory of a quantum equilibrium ensemble. This is usually trivial in
pilot-wave theory, where the velocity field (\ref{v}) is equal to the
equilibrium probability current divided by the equilibrium probability
density. But in the case of the Wheeler-DeWitt equation (\ref{WD}) there is no
generally well-behaved candidate for either of these quantities. If one
attempts to straightforwardly interpret $\left\vert \Psi\lbrack g_{ij}%
]\right\vert ^{2}$ as a probability density for quantum equilibrium, one is
left with the difficulty of recovering a time dependence at the quantum level
(where in general in de Broglie-Bohm theory the details of the trajectories
are not observable in equilibrium). A common approach to solving this problem
is to extract an appropriate degree of freedom $\tilde{t}[g_{ij}]$ from the
3-metric to play the role of time, so that $\Psi\lbrack g_{ij}]$ effectively
becomes of the schematic form $\Psi\lbrack\tilde{g}_{ij},\tilde{t}]$ where
$\tilde{g}_{ij}$ are the remaining metric variables. In quantum cosmology, for
example, a popular choice of time variable is the scale factor $a$ for an
expanding universe. However, while this method certainly works in some cases,
there seems to be no generally consistent way of extracting a well-behaved
time function \cite{UW89,Ish91,Kuch11}. (For example, for a closed universe
the `time' $a$ appears to stop and reverse at the point of maximum expansion,
making it difficult to ensure that only one physical state is associated with
each value of time.) As a result, there seems to be no generally well-behaved
equilibrium current or time-dependent density for appropriate degrees of
freedom $\tilde{g}_{ij}$.

Here we are touching on the notoriously controversial `problem of time' in
canonical quantum gravity. On one viewpoint, it might be asserted that our
usual notion of time is meaningful only in certain emergent regimes, in which
case it is to be expected that there is no generally well-defined time
evolution. On another viewpoint, it might be suggested that the formalism is
afflicted with a serious conceptual difficulty.\footnote{Loop quantum gravity
\cite{Rov04,Thiem08,Boj11,GamPu11} has technical advantages over the older
metric representation being used here, but does not significantly improve the
conceptual problem of time.}

On either view, a well-behaved equilibrium current and density may be
generally said to emerge in those regimes where there is a
Schr\"{o}dinger-like equation $i\partial\Psi/\partial\tilde{t}=\widehat
{\tilde{H}}\Psi$ for a wave functional $\Psi\lbrack\tilde{g}_{ij},\tilde{t}]$,
with an effective Hamiltonian $\widehat{\tilde{H}}$ and time parameter
$\tilde{t}$. If $\widehat{\tilde{H}}$ is given by a differential operator
there will be an associated continuity equation \cite{SV09}%
\begin{equation}
\frac{\partial|\Psi|^{2}}{\partial\tilde{t}}+\int d^{3}x\ \frac{\delta J_{ij}%
}{\delta\tilde{g}_{ij}}=0\ ,
\end{equation}
where $J_{ij}$ is a current. We may then define a de Broglie velocity field
$\partial\tilde{g}_{ij}/\partial\tilde{t}=J_{ij}/|\Psi|^{2}$ and an ensemble
of 3-geometries with an arbitrary distribution $P$ of metrics $\tilde{g}_{ij}$
will evolve according to%
\[
\frac{\partial P}{\partial\tilde{t}}+\int d^{3}x\ \frac{\delta}{\delta
\tilde{g}_{ij}}\left(  P\frac{\partial\tilde{g}_{ij}}{\partial\tilde{t}%
}\right)  =0\ .
\]
An ensemble with a distribution $P=|\Psi|^{2}$ at some initial time will then
evolve into an ensemble with a distribution $P=|\Psi|^{2}$ at later times --
the system will possess a quantum equilibrium state.

Outside of this `Schr\"{o}dinger regime', however, there appears to be no good
reason to expect a quantum equilibrium state to exist. Of course, the mere
fact that one cannot apply the usual derivation of an equilibrium state does
not by itself imply that there is no such state. But it is suggestive. It may
simply be a mistake to assume that quantum gravity generally possesses a
quantum equilibrium state described by a Born-like rule. We propose, then,
that quantum equilibrium exists only in the Schr\"{o}dinger-like regime. In
pilot-wave theory, which is ultimately a dynamics of individual systems and
not a dynamics of ensembles, it is in principle possible to consider this
proposal in a conceptually coherent manner.

We may then expect quantum nonequilibrium to be generated by
quantum-gravitational processes at the Planck scale. Schematically, consider
an incoming state $\mathcal{S}_{\mathrm{in}}$ (with a wave functional and de
Broglie-Bohm 3-geometry) that is accurately described by a Schr\"{o}dinger
regime, with a Schr\"{o}dinger equation $i\partial\Psi_{\mathrm{in}}%
/\partial\tilde{t}_{\mathrm{in}}=\widehat{\tilde{H}}_{\mathrm{in}}%
\Psi_{\mathrm{in}}$ and a quantum equilibrium state $\left\vert \Psi
_{\mathrm{in}}\right\vert ^{2}$. Let us assume that the incoming state is
indeed in equilibrium. The state could subsequently encounter interactions in
the deep quantum-gravity regime, for which there is no well-defined
Schr\"{o}dinger equation and no well-defined quantum equilibrium state (a
`non-Schr\"{o}dinger regime'). One may end with an outgoing state
$\mathcal{S}_{\mathrm{out}}$ that is again accurately described by a
Schr\"{o}dinger regime. However, the outgoing Schr\"{o}dinger equation
$i\partial\Psi_{\mathrm{out}}/\partial\tilde{t}_{\mathrm{out}}=\widehat
{\tilde{H}}_{\mathrm{out}}\Psi_{\mathrm{out}}$ and the outgoing quantum
equilibrium state $\left\vert \Psi_{\mathrm{out}}\right\vert ^{2}$ may or may
not coincide with their ingoing counterparts. In the absence of a single
Schr\"{o}dinger equation and associated probability current that describes the
entire evolution from $\mathcal{S}_{\mathrm{in}}$ to $\mathcal{S}%
_{\mathrm{out}}$, it is not possible to prove that an initial equilibrium
state evolves to a final equilibrium state by integrating a single continuity
equation. In such circumstances, there seems to be no obstruction to an
incoming equilibrium state evolving into an outgoing \textit{non}equilibrium
state. Such a transition, from equilibrium to nonequilibrium, could be
established only in the context of a specific model. For example, such a
scenario might be naturally applied to a bouncing model of quantum cosmology
\cite{PPreview,PNAVinprep}. More generally, our arguments suggest that quantum
nonequilibrium could be generated for processes taking place at the Planck
scale -- such as the exit of trans-Planckian field modes from the Planck
radius during inflation.

\subsection{Regularisation of phase singularities in pilot-wave dynamics}

Like classical general relativity, pilot-wave dynamics predicts its own
demise. For standard Hamiltonians that are quadratic in the canonical momenta,
the velocity field (\ref{vst}) generally diverges at nodes (where $\psi=0$).
Nodes are also known as `phase singularities', where the phase
$S=\operatorname{Im}\ln\psi$ becomes ill-defined \cite{Berry81}. In a general
$n$-dimensional configuration space, nodes form $(n-2)$-dimensional surfaces
(as is clear from consideration of the simultaneous equations
$\operatorname{Re}\psi=0$, $\operatorname{Im}\psi=0$ at fixed time $t$).

Thus for standard Hamiltonians pilot-wave theory breaks down at nodes. This
elementary point is usually disregarded. In practice the divergence can be
ignored because nodes form a set of measure zero (with respect to the standard
volume measure in configuration space). Even so, as a matter of principle the
dynamics breaks down in these regions, signalling the possibility of new
physics there.

While this divergence afflicts systems with Hamiltonians that are quadratic in
the canonical momenta -- for example nonrelativistic spinless particles -- for
some well-known systems the Hamiltonian is not of that form and there is no
divergence. In particular, for a (high-energy)\ Dirac electron the one-body
Dirac equation for a 4-component spinor $\psi$ has a conserved current density
$j^{\mu}=(j^{0},j^{i})=(\bar{\psi}\gamma^{0}\psi,\bar{\psi}\gamma^{i}\psi)$
(where the $\gamma^{\mu}$ are Dirac matrices) which may be used to define a
natural velocity field $v^{i}=j^{i}/j^{0}$ and a de Broglie guidance equation
$dx^{i}/dt=\bar{\psi}\gamma^{i}\psi/\bar{\psi}\gamma^{0}\psi$ \cite{Holl93}.
(A similar construction may be given for the many-body case
\cite{BH93,C03,CS07}.) This velocity field is finite everywhere and indeed
bounded by the speed of light $c$. It might then be suspected that the
divergences could be an artifact of the low-energy, nonrelativistic theory.
However, divergence at nodes is found in high-energy bosonic field theory,
just as in the nonrelativistic particle case. For example, for a single
(unentangled) mode $\mathbf{k}$ of a free massless and real scalar field
$\phi$ on Minkowski spacetime, if we write the Fourier components in terms of
their real and imaginary parts, $\phi_{\mathbf{k}}\propto\left(
q_{\mathbf{k}1}+iq_{\mathbf{k}2}\right)  $ (cf. Section 3), the wave function
$\psi_{\mathbf{k}}=\psi_{\mathbf{k}}(q_{\mathbf{k}1},q_{\mathbf{k}2},t)$ of
the mode satisfies \cite{AV07,AV08}%
\[
i\frac{\partial\psi_{\mathbf{k}}}{\partial t}=-\frac{1}{2}\left(
\frac{\partial^{2}}{\partial q_{\mathbf{k}1}^{2}}+\frac{\partial^{2}}{\partial
q_{\mathbf{k}2}^{2}}\right)  \psi_{\mathbf{k}}+\frac{1}{2}k^{2}\left(
q_{\mathbf{k}1}^{2}+q_{\mathbf{k}2}^{2}\right)  \psi_{\mathbf{k}}\ ,
\]
and the de Broglie velocities for $q_{\mathbf{k}r}$ ($r=1,2$) are%
\[
\frac{dq_{\mathbf{k}r}}{dt}=\frac{\partial s_{\mathbf{k}}}{\partial
q_{\mathbf{k}r}}%
\]
(with $\psi_{\mathbf{k}}=\left\vert \psi_{\mathbf{k}}\right\vert
e^{is_{\mathbf{k}}}$). These equations are the same as in the pilot-wave
theory of a nonrelativistic particle of unit mass in a simple harmonic
oscillator potential in the $q_{\mathbf{k}1}-q_{\mathbf{k}2}$ plane, and the
velocity field $\dot{q}_{\mathbf{k}r}$ exhibits the same divergence at nodes.
Therefore, the physical motivation remains even in high-energy field theory.

If pilot-wave theory is taken seriously as a physical theory, the divergences
must be removed or regularised by some mechanism. We suggest that their
presence may be taken as a sign that new physics is needed at very short
distances in configuration space (just as the presence of singularities in
general relativity signals the need for new physics at very short distances in
spacetime). Such new physics will presumably result in corrections to quantum
mechanics. To find this new physics, one approach would be to seek new
fundamental principles. Alternatively (or concurrently), one may develop
simple phenomenological models and attempt to constrain them experimentally.
The latter approach is followed here.

The need for regularisation in pilot-wave theory was briefly recognised in a
paper by Bell (ref. \cite{Bell87}, p. 138) where in a footnote it was remarked
that the velocity field (\ref{v}) may be regularised by smearing the numerator
$j$ and denominator $|\psi|^{2}$ with a narrowly-peaked function (Bell
suggested a Gaussian) in such a way that the smeared $|\psi|^{2}$ becomes the
new equilibrium distribution.

While Bell did not write down any equations, his intentions are clear and
easily reconstructed. Introducing a narrowly-peaked and positive-definite
weighting function $\mu(q^{\prime}-q)$ (for example a Gaussian) on
configuration space, where $\int dq^{\prime}\ \mu(q^{\prime}-q)=1$, we may
define a regularised current%
\begin{equation}
j(q,t)_{\mathrm{reg}}=\int dq^{\prime}\ \mu(q^{\prime}-q)j(q^{\prime},t)\ ,
\label{jreg}%
\end{equation}
a regularised density%
\begin{equation}
\left(  |\psi(q,t)|^{2}\right)  _{\mathrm{reg}}=\int dq^{\prime}%
\ \mu(q^{\prime}-q)|\psi(q^{\prime},t)|{^{2}}\ , \label{psi2reg}%
\end{equation}
and a regularised velocity field%
\begin{equation}
v(q,t)_{\mathrm{reg}}=\frac{j(q,t)_{\mathrm{reg}}}{(|\psi(q,t)|^{2}%
)_{\mathrm{reg}}}\ . \label{vreg}%
\end{equation}
The latter may also be written as a `mean'%
\begin{equation}
v(q,t)_{\mathrm{reg}}=\frac{\int dq^{\prime}\ \mu(q^{\prime}-q)|\psi
(q^{\prime},t)|^{2}v(q^{\prime},t)}{\int dq^{\prime}\ \mu(q^{\prime}%
-q)|\psi(q^{\prime},t)|^{2}} \label{vreg'}%
\end{equation}
of the unregularised field $v$ with a weighting function $\mu|\psi|^{2}$.

We may then adopt the modified de Broglie equation of motion for the
trajectories,%
\begin{equation}
\frac{dq}{dt}=v(q,t)_{\mathrm{reg}}\ , \label{deBgenreg}%
\end{equation}
together with the usual Schr\"{o}dinger equation (\ref{Sch1}) for $\psi$.
Equations (\ref{deBgenreg}) and (\ref{Sch1}) may be taken as the basic
equations of a regularised pilot-wave dynamics. Assuming that $|\psi|^{2}$
vanishes only in regions of zero Lebesgue measure, $(|\psi|^{2})_{\mathrm{reg}%
}$ will be positive everywhere and the new velocity field $v_{\mathrm{reg}}$
will indeed be regular everywhere. The unregularised theory is recovered as
the smearing function $\mu$ becomes arbitrarily narrow, $\mu(q^{\prime
}-q)\rightarrow\delta(q^{\prime}-q)$.

Using (\ref{Contj}), together with $\partial_{q}\mu(q^{\prime}-q)=-\partial
_{q}^{\prime}\mu(q^{\prime}-q)$ (where we write $\partial_{q}^{\prime
}=\partial_{q^{\prime}}$), one finds that%
\begin{equation}
\frac{\partial(|\psi|^{2})_{\mathrm{reg}}}{\partial t}+\partial_{q}\cdot
j_{\mathrm{reg}}=0
\end{equation}
or%
\begin{equation}
\frac{\partial(|\psi|^{2})_{\mathrm{reg}}}{\partial t}+\partial_{q}%
\cdot\left(  (|\psi|^{2})_{\mathrm{reg}}v_{\mathrm{reg}}\right)
=0\ .\label{contpsi2reg}%
\end{equation}
If we again consider an ensemble of systems with the same wave function
$\psi(q,t)$, the time evolution of an arbitrary distribution $\rho(q,t)$ of
configurations $q(t)$ will now obey the regularised continuity equation%
\begin{equation}
\frac{\partial\rho}{\partial t}+\partial_{q}\cdot(\rho v_{\mathrm{reg}%
})=0\ .\label{contrhoreg}%
\end{equation}
Comparison of (\ref{contpsi2reg}) and (\ref{contrhoreg}) shows that an initial
distribution $\rho(q,t_{i})=(|\psi(q,t_{i})|^{2})_{\mathrm{reg}}$ at time
$t_{i}$ evolves into a final distribution%
\begin{equation}
\rho(q,t)=(|\psi(q,t)|^{2})_{\mathrm{reg}}\label{qereg}%
\end{equation}
at time $t$ (where in general $(|\psi|^{2})_{\mathrm{reg}}\neq|\psi|^{2}$).

Thus the quantum equilibrium state is modified, or smeared, by the (narrow)
regulator function $\mu$, inducing deviations from the Born rule at small
lengthscales in configuration space. As in the unregularised theory, we may
expect to find relaxation $\rho(q,t)\rightarrow(|\psi(q,t)|^{2})_{\mathrm{reg}%
}$ as quantified by the decrease of an $H$-function $H_{\mathrm{reg}}=\int
dq\ \rho\ln(\rho/(|\psi|^{2})_{\mathrm{reg}})$ (on a coarse-grained level).

The regularised equations are not supposed to be a candidate for a fundamental
theory but instead are expected to provide an effective description of some
deeper physics taking place at very short distances (much as in the analogous
case of quantum field theory).

As we discuss in detail elsewhere, generally speaking it would be worth
conducting experiments to probe the quantum probability distribution on small
scales and to search for deviations from the Born rule in regions where the
standard quantum-theoretical probability density approaches zero
\cite{AVnodesprep,AVbook}. In this paper we focus on the possible relevance to
inflationary cosmology and physics at the Planck scale.

We have said that the regulator function $\mu$ should be regarded as an
effective description of new physics at short distances in configuration
space. In the above construction we assumed that $\mu$ was independent of
time. But if we consider inflationary field modes that evolve from
sub-Planckian to super-Planckian physical wavelengths, then because the modes
make a transition between such different physical regimes it is plausible to
suppose that during the transition $\mu$ could be time dependent. As we shall
see, if the regulator function $\mu$ depends on time as the mode exits the
Planck radius then quantum nonequilibrium will be generated from a prior
equilibrium state -- that is, the Born rule will become unstable at the Planck scale.

\section{Pilot-wave dynamics and inflation}

In Section 2.1 we formulated the pilot-wave theory of a massless (and
minimally-coupled) real scalar field $\phi$ on a general globally-hyperbolic
spacetime, with an assumed preferred foliation. Let us now consider the same
field on an expanding flat space, with spacetime line element $d\tau
^{2}=dt^{2}-a^{2}d\mathbf{x}^{2}$ (where $a=a(t)$ is the scale factor and we
take $c=1$). This corresponds to a case with a uniform lapse function $N=1$
and a 3-metric $g_{ij}=a^{2}\delta_{ij}$ (with $g=a^{6}$). The dynamical
equations (\ref{Sch2}) and (\ref{deB2}) then become%
\begin{equation}
i\frac{\partial\Psi}{\partial t}=\int d^{3}\mathbf{x}\;\left(  -\frac
{1}{2a^{3}}\frac{\delta^{2}}{\delta\phi^{2}}+\frac{1}{2}a(\mathbf{\nabla}%
\phi)^{2}\right)  \Psi
\end{equation}
and%
\begin{equation}
\frac{\partial\phi}{\partial t}=\frac{1}{a^{3}}\frac{\delta S}{\delta\phi}\ .
\end{equation}
Working with Fourier components $\phi_{\mathbf{k}}=\frac{\sqrt{V}}%
{(2\pi)^{3/2}}\left(  q_{\mathbf{k}1}+iq_{\mathbf{k}2}\right)  $ -- with $V$ a
normalisation volume and $q_{\mathbf{k}r}$ ($r=1,2$) real variables -- we have
a Schr\"{o}dinger equation%
\begin{equation}
i\frac{\partial\Psi}{\partial t}=\sum_{\mathbf{k}r}\left(  -\frac{1}{2a^{3}%
}\frac{\partial^{2}}{\partial q_{\mathbf{k}r}^{2}}+\frac{1}{2}ak^{2}%
q_{\mathbf{k}r}^{2}\right)  \Psi\label{Sch3}%
\end{equation}
for $\Psi=\Psi\lbrack q_{\mathbf{k}r},t]$ and de Broglie velocities%
\begin{equation}
\frac{dq_{\mathbf{k}r}}{dt}=\frac{1}{a^{3}}\operatorname{Im}\frac{1}{\Psi
}\frac{\partial\Psi}{\partial q_{\mathbf{k}r}}=\frac{1}{a^{3}}\frac{\partial
S}{\partial q_{\mathbf{k}r}} \label{deB3}%
\end{equation}
(with $\Psi=\left\vert \Psi\right\vert e^{iS}$) for the evolving degrees of
freedom $q_{\mathbf{k}r}$ \cite{AV07,AV08,AV10}. The time evolution of an
arbitrary distribution $P[q_{\mathbf{k}r},t]$ is given by%
\begin{equation}
\frac{\partial P}{\partial t}+\sum_{\mathbf{k}r}\frac{\partial}{\partial
q_{\mathbf{k}r}}\left(  P\frac{1}{a^{3}}\frac{\partial S}{\partial
q_{\mathbf{k}r}}\right)  =0\ . \label{cont3}%
\end{equation}

An unentangled mode $\mathbf{k}$ has an independent dynamics with two degrees
of freedom $q_{\mathbf{k}1}$, $q_{\mathbf{k}2}$. This has been used
extensively to study cosmological relaxation for a radiation-dominated
expansion with $a\propto t^{1/2}$ \cite{AV07,AV08,AV10,CV13,CV14}. In the
short-wavelength or sub-Hubble limit, we obtain the time evolution of a field
mode on Minkowski spacetime and rapid relaxation takes place for a
superposition of excited states; whereas for long (super-Hubble) wavelengths
it is found that relaxation is retarded. If there was a radiation-dominated
pre-inflationary era, we may expect incomplete relaxation at sufficiently long
wavelengths -- resulting in a large-scale power deficit in the inflationary
spectrum \cite{AV10,CV13,CV14}.

Incomplete relaxation during a pre-inflationary era is one means by which
nonequilibrium could exist in the inflationary spectrum. Another possibility
-- the subject of this paper -- is that nonequilibrium is generated during
inflation itself by novel gravitational effects at the Planck scale. This was
suggested in ref. \cite{AV10} (section IVB). As we have noted, trans-Planckian
modes -- that is, modes that originally had sub-Planckian physical wavelengths
$\lambda_{\mathrm{phys}}=a\lambda=a(2\pi/k)$ -- may well make an observable
contribution to the inflationary spectrum \cite{BM01,MB01}, in which case
inflation will allow us to probe physics at the Planck scale \cite{BM13}. If
quantum nonequilibrium is indeed generated at the Planck length $l_{\mathrm{P}%
}$, an equilibrium mode with a physical wavelength $\lambda_{\mathrm{phys}%
}<l_{\mathrm{P}}$ in the early inflationary era would be driven out of
equilibrium upon exiting the Planck radius (that is, when $\lambda
_{\mathrm{phys}}>l_{\mathrm{P}}$) \cite{AV10}. The inflaton field would then
carry quantum nonequilibrium at short wavelengths (below a comoving cutoff).

However, a specific model of such a process has yet to be constructed. We
shall do so here (Section 4). Our model of quantum instability will employ the
results of refs. \cite{AV07,AV10}, in which we calculated the de Broglie-Bohm
trajectories for the inflaton perturbation $\phi$ in the inflationary
(Bunch-Davies) vacuum. We recall the key results that will be needed to
construct our model.

It is convenient to use conformal time $\eta$ defined by $d\eta=dt/a$. For
$a\propto e^{Ht}$ we have $\eta=-1/Ha$ and on an idealised de Sitter space
$\eta$ ranges over $(-\infty,0)$. The Bunch-Davies vacuum wave functional
$\Psi\lbrack q_{\mathbf{k}r},\eta]$ is a product $\prod\limits_{\mathbf{k}%
r}\psi_{\mathbf{k}r}(q_{\mathbf{k}r},\eta)$ of contracting Gaussian packets
$\psi_{\mathbf{k}r}(q_{\mathbf{k}r},\eta)$. Considering a single degree of
freedom $q_{\mathbf{k}r}$ we may drop the index $\mathbf{k}r$. Writing the
wave function $\psi=\psi(q,\eta)$ as $\psi=\left\vert \psi\right\vert e^{is}$,
the conformal de Broglie velocity for the trajectory $q=q(\eta)$ is given by
$dq/d\eta=adq/dt$ or (using (\ref{deB3}))%
\begin{equation}
\frac{dq}{d\eta}=(H\eta)^{2}\frac{\partial s}{\partial q}\ . \label{deBconf}%
\end{equation}
As shown in ref. \cite{AV10}, the squared amplitude of $\psi$ is a Gaussian%
\begin{equation}
\left\vert \psi(q,\eta)\right\vert ^{2}=\frac{1}{\sqrt{2\pi\Delta^{2}}%
}e^{-q^{2}/2\Delta^{2}} \label{psi2}%
\end{equation}
with a contracting width%
\begin{equation}
\Delta(\eta)=\Delta(0)\sqrt{1+k^{2}\eta^{2}}\ , \label{Delta}%
\end{equation}
where for convenience we write in terms of the asymptotic value%
\begin{equation}
\Delta(0)=H/\sqrt{2k^{3}}\ .
\end{equation}
For a calculation over a finite time interval $(\eta_{i},\eta_{f})$, we may
write%
\begin{equation}
\Delta(0)=\Delta(\eta_{i})/\sqrt{1+k^{2}\eta_{i}^{2}}\ .
\end{equation}
The phase of $\psi$ is given by%
\begin{equation}
s(q,\eta)=\frac{1}{2H^{2}}\frac{q^{2}}{\eta(\eta^{2}+1/k^{2})}+h(\eta)
\label{s}%
\end{equation}
where $h(\eta)=\frac{1}{2}\left(  -k\eta+\tan^{-1}\left(  k\eta\right)
\right)  $ is independent of $q$. Thus from (\ref{deBconf}) we have%
\begin{equation}
\frac{dq}{d\eta}=\frac{q\eta}{\eta^{2}+1/k^{2}}\ . \label{deBode}%
\end{equation}
The trajectories then take the simple form%
\[
q(\eta)=q(0)\sqrt{1+k^{2}\eta^{2}}%
\]
(where again for convenience we write in terms of the asymptotic value
$q(0)=q(\eta_{i})/\sqrt{1+k^{2}\eta_{i}^{2}}$).

The time evolution of an arbitrary distribution $\rho(q,\eta)$ is given by the
general solution%
\begin{equation}
\rho(q,\eta)=\frac{1}{\sqrt{1+k^{2}\eta^{2}}}\rho(q/\sqrt{1+k^{2}\eta^{2}},0)
\end{equation}
of the continuity equation%
\begin{equation}
\frac{\partial\rho}{\partial\eta}+\frac{\partial}{\partial q}\left(  \rho
q^{\prime}\right)  =0
\end{equation}
(where $q^{\prime}\equiv dq/d\eta$). The distribution has a contracting width%
\begin{equation}
D(\eta)=D(0)\sqrt{1+k^{2}\eta^{2}}\ .
\end{equation}

We obtain a homogeneous contraction of both $\rho$ and $\left\vert
\psi\right\vert ^{2}$ by the same rescaling factor $1/\sqrt{1+k^{2}\eta^{2}}$.
Thus for each degree of freedom $q_{\mathbf{k}r}$ the width of $\rho$ remains
in a constant ratio with the width of $\left\vert \psi\right\vert ^{2}$.

The ratio%
\begin{equation}
\xi(k)\equiv\frac{\left\langle |\phi_{\mathbf{k}}|^{2}\right\rangle
}{\left\langle |\phi_{\mathbf{k}}|^{2}\right\rangle _{\mathrm{QT}}}
\label{ksi}%
\end{equation}
of the nonequilibrium variance $\left\langle |\phi_{\mathbf{k}}|^{2}%
\right\rangle $ to the quantum-theoretical variance $\left\langle
|\phi_{\mathbf{k}}|^{2}\right\rangle _{\mathrm{QT}}$ is then preserved in
time. Relic nonequilibrium ($\xi\neq1$) at the beginning of inflation will be
preserved during the inflationary era and transferred to larger physical
wavelengths $\lambda_{\mathrm{phys}}$ by the spatial expansion. By the same
token, of course, initial equilibrium ($\xi=1$) is also preserved in time. As
it stands, the model does not allow nonequilibrium to be created from a prior
equilibrium state. We shall now consider a modification of pilot-wave dynamics
in which this is possible.

\section{A model for quantum instability}

In Section 2.3 we discussed the regularisation of phase singularities in
pilot-wave theory. By smearing the quantum density and current with a narrow
function $\mu(q^{\prime}-q)$ we may define a regularised velocity field
(\ref{vreg}). This yields a modified equilibrium state (\ref{qereg}) with
deviations from the Born rule at small distances in configuration space. These
deviations become arbitrarily small as $\mu(q^{\prime}-q)\rightarrow
\delta(q^{\prime}-q)$. However, this construction assumes that the regulator
function $\mu$ has no time dependence. But if $\mu$ is an effective
description of new physics at short distances, and if we consider inflationary
field modes that transition from the sub-Planckian regime ($\lambda
_{\mathrm{phys}}<l_{\mathrm{P}}$) to the super-Planckian regime ($\lambda
_{\mathrm{phys}}>l_{\mathrm{P}}$), then it appears reasonable to allow $\mu$
to be time dependent during the transition. As we shall now show, quantum
nonequilibrium can then be generated from a prior equilibrium state. For a
given regulator $\mu=\mu(q^{\prime}-q,t)$ it is possible to calculate the time
evolution away from equilibrium as the mode exits the Planck radius and so
obtain an expression for the function $\xi(k)$ which quantifies deviations
from the Born rule in the inflationary power spectrum.

\subsection{Creation of nonequilibrium by a time-dependent regulator}

For a general system with configuration $q$ and time-dependent regulator
$\mu(q^{\prime}-q,t)$ (again with $\int dq^{\prime}\ \mu(q^{\prime}-q,t)=1$)
we may still define a regularised current%
\begin{equation}
j(q,t)_{\mathrm{reg}}=\int dq^{\prime}\ \mu(q^{\prime}-q,t)j(q^{\prime
},t)\label{tregj}%
\end{equation}
and a regularised density%
\begin{equation}
\left(  |\psi(q,t)|^{2}\right)  _{\mathrm{reg}}=\int dq^{\prime}%
\ \mu(q^{\prime}-q,t)|\psi(q^{\prime},t)|{^{2}}\ ,\label{tregpsi2}%
\end{equation}
with a regularised velocity field $v(q,t)_{\mathrm{reg}}=j(q,t)_{\mathrm{reg}%
}/(|\psi(q,t)|^{2})_{\mathrm{reg}}$ or%
\begin{equation}
v(q,t)_{\mathrm{reg}}=\frac{\int dq^{\prime}\ \mu(q^{\prime}-q,t)|\psi
(q^{\prime},t)|^{2}v(q^{\prime},t)}{\int dq^{\prime}\ \mu(q^{\prime}%
-q,t)|\psi(q^{\prime},t)|^{2}}%
\end{equation}
as before. We still have the de Broglie equation of motion
$dq/dt=v(q,t)_{\mathrm{reg}}$ for the trajectories and the Schr\"{o}dinger
equation (\ref{Sch1}) for $\psi$.

The time dependence of $\mu(q^{\prime}-q,t)$ does however make one crucial
difference: the regularised density $\left(  |\psi(q,t)|^{2}\right)
_{\mathrm{reg}}$ is no longer an equilibrium state. To see this, note that an
arbitrary distribution $\rho(q,t)$ still obeys the regularised continuity
equation (\ref{contrhoreg}) whereas $\left(  |\psi(q,t)|^{2}\right)
_{\mathrm{reg}}$ no longer obeys the (same) continuity equation
(\ref{contpsi2reg}). Instead, $\left(  |\psi(q,t)|^{2}\right)  _{\mathrm{reg}%
}$ satisfies%
\begin{equation}
\frac{\partial(|\psi|^{2})_{\mathrm{reg}}}{\partial t}+\partial_{q}%
\cdot\left(  (|\psi|^{2})_{\mathrm{reg}}v_{\mathrm{reg}}\right)
=s\label{contsource}%
\end{equation}
where the `source term' $s$ is given by%
\begin{equation}
s(q,t)=\int dq^{\prime}\ \frac{\partial\mu(q^{\prime}-q,t)}{\partial t}%
|\psi(q^{\prime},t)|^{2}\ .\label{sdefn}%
\end{equation}
(This follows from (\ref{Contj}) with $\partial_{q}\mu(q^{\prime}%
-q)=-\partial_{q}^{\prime}\mu(q^{\prime}-q)$.)

It now follows that an initial distribution $\rho(q,t_{i})=(|\psi
(q,t_{i})|^{2})_{\mathrm{reg}}$ at time $t_{i}$ in general evolves into a
final distribution%
\begin{equation}
\rho(q,t)\neq(|\psi(q,t)|^{2})_{\mathrm{reg}}%
\end{equation}
at time $t$, so that indeed $\left(  |\psi(q,t)|^{2}\right)  _{\mathrm{reg}}$
is not an equilibrium state. This will be confirmed below for a simple
example. In general, from (\ref{contrhoreg}) and (\ref{contsource}) it follows
that the regularised ratio%
\[
f_{\mathrm{reg}}\equiv\frac{\rho}{(|\psi|^{2})_{\mathrm{reg}}}%
\]
satisfies%
\begin{equation}
\frac{df_{\mathrm{reg}}}{dt}=-uf_{\mathrm{reg}} \label{dfreg}%
\end{equation}
where%
\[
u\equiv\frac{s}{(|\psi|^{2})_{\mathrm{reg}}}%
\]
and now $d/dt=\partial/\partial t+v_{\mathrm{reg}}\cdot\partial_{q}$.
Integrating (\ref{dfreg}) along a trajectory from an initial point $q_{i}$ at
time $t_{i}$ to a final point $q_{f}$ at time $t_{f}$ we have%
\[
f_{\mathrm{reg}}(q_{f},t_{f})=f_{\mathrm{reg}}(q_{i},t_{i}).\exp\left(
-\int_{\mathrm{traj}}dt\ u(q(t),t)\right)  \ .
\]
In general, $\int_{\mathrm{traj}}udt\neq0$ and $f_{\mathrm{reg}}(q_{f}%
,t_{f})\neq f_{\mathrm{reg}}(q_{i},t_{i})$.

We may then consider the following type of scenario. At times $t<t_{i}$ and
$t>t_{f}$ the regulator $\mu$ is time independent and $s=0$. At these times we
will have a regularised equilibrium distribution $(|\psi|^{2})_{\mathrm{reg}}%
$. Should $\mu$ be time dependent during the interval $(t_{i},t_{f})$, then an
incoming equilibrium state $\rho=(|\psi|^{2})_{\mathrm{reg}}$ will evolve into
an outgoing \textit{non}equilibrium state $\rho\neq(|\psi|^{2})_{\mathrm{reg}%
}$.

Such a scenario may be applied to a case where for $t<t_{i}$ and $t>t_{f}$ the
regularisation may be neglected, with $\mu(q^{\prime}-q,t)=\delta(q^{\prime
}-q)$ (approximately). At these times we will have the standard Born-rule
equilibrium distribution $(|\psi|^{2})_{\mathrm{reg}}=$ $|\psi|^{2}$. If $\mu$
is time dependent during the interval $(t_{i},t_{f})$, an incoming Born-rule
distribution $\rho=|\psi|^{2}$ will evolve into an outgoing non-Born-rule
distribution $\rho\neq|\psi|^{2}$.

\subsection{Calculation of the nonequilibrium function $\xi(k)$}

We may now apply these considerations to an inflationary field mode, yielding
a model in which quantum nonequilibrium is created during inflation as
trans-Planckian modes exit the Planck radius. For this purpose it will be
convenient to use conformal time $\eta$. As we saw in Section 3, the
Bunch-Davies vacuum wave function $\psi(q,\eta)$ is a contracting Gaussian.
While this wave function does not possess nodes ($\psi\neq0$ for all finite
$q$), even so the generic presence of nodes for arbitrary wave functions --
generally superpositions of the vacuum with excited states -- implies a need
for regularisation at short distances in configuration space. As we discussed
in Section 2.3, such regularisation may be viewed as an effective description
of new physics. We will assume this new physics to be present even if $\psi$
happens to be free of nodes.

We may now reconsider the results for the inflationary vacuum (summarised in
Section 3) including the presence of a time-dependent regulator $\mu
(q^{\prime}-q,\eta)$. For definiteness we consider a simple example. We take%
\begin{equation}
\mu(q,\eta)=\delta_{\alpha}(q)=\frac{1}{\sqrt{2\pi\alpha^{2}}}e^{-q^{2}%
/2\alpha^{2}}\ ,\label{finitedelta}%
\end{equation}
where $\delta_{\alpha}(q)$ is a regularised delta-function of time-dependent
width $\alpha=\alpha(\eta)$ (with $\alpha\geq0$ for all $\eta$). Note that
$\delta_{\alpha}(q)\rightarrow\delta(q)$ as $\alpha\rightarrow0$. We take%
\begin{equation}
\alpha(\eta_{i})=\alpha(\eta_{f})=0\ ,\label{alphaif}%
\end{equation}
so that the regulator is switched off at the initial and final times $\eta
_{i}$ and $\eta_{f}$. Strictly speaking, it would be more realistic to take
$\alpha(\eta_{i})$ and $\alpha(\eta_{f})$ to be very small but non-zero, so
that the regulator is negligible initially and finally. We assume that
$\alpha=\alpha(\eta)$ is non-negligible and time dependent during the interval
$(\eta_{i},\eta_{f})$, so that an incoming equilibrium state evolves into an
outgoing nonequilibrium state.

We assume that the interval $(\eta_{i},\eta_{f})$ straddles the time at which
the mode exits the Planck radius. Thus, according to this model,
regularisation is constant and negligible in the far sub-Planckian and far
super-Planckian regimes but regularisation is non-negligible and time
dependent during Planck radius crossing. As a consequence, our model then
describes the creation of nonequilibrium for trans-Planckian modes emerging
from the Planck radius.

Applying the Gaussian regulator (\ref{finitedelta}) to the Bunch-Davies
vacuum, from (\ref{tregpsi2}) we find a regularised density%
\begin{equation}
(|\psi|^{2})_{\mathrm{reg}}=\frac{1}{\sqrt{2\pi(\Delta^{2}+\alpha^{2})}%
}e^{-q^{2}/2(\Delta^{2}+\alpha^{2})}\ .
\end{equation}
This is still a Gaussian packet but with a modified width $\sqrt{\Delta
^{2}+\alpha^{2}}$. From (\ref{tregj}) we also find a regularised current%
\begin{equation}
j_{\mathrm{reg}}=\frac{\Delta^{2}}{\Delta^{2}+\alpha^{2}}\frac{q\eta}{\eta
^{2}+1/k^{2}}(|\psi|^{2})_{\mathrm{reg}}\ .
\end{equation}
The regularised de Broglie velocity is then given by%
\begin{equation}
v(q,\eta)_{\mathrm{reg}}\equiv j_{\mathrm{reg}}/(|\psi|^{2})_{\mathrm{reg}%
}=\frac{\Delta^{2}}{\Delta^{2}+\alpha^{2}}\frac{q\eta}{\eta^{2}+1/k^{2}}%
=\frac{\Delta^{2}}{\Delta^{2}+\alpha^{2}}v(q,\eta)\ ,
\end{equation}
where $v(q,\eta)$ is the unregularised velocity field (\ref{deBode}). It is
convenient to define a function $g^{2}(\eta)$ by%
\begin{equation}
g^{2}=(1/k^{2})(1+\alpha^{2}/\Delta_{0}^{2})\ ,\label{g2}%
\end{equation}
where (from (\ref{Delta})) $\Delta_{0}^{2}=\Delta^{2}/(1+k^{2}\eta^{2})$ is
the unregularised equilibrium variance at $\eta=0$. From (\ref{alphaif}) we
have $g^{2}(\eta_{i})=g^{2}(\eta_{f})=1/k^{2}$. Since%

\begin{equation}
\frac{\Delta^{2}}{\Delta^{2}+\alpha^{2}}=\frac{\Delta_{0}^{2}(1+k^{2}\eta
^{2})}{\Delta_{0}^{2}(1+k^{2}\eta^{2})+\alpha^{2}}=\frac{\eta^{2}+1/k^{2}%
}{\eta^{2}+g^{2}}\ ,
\end{equation}
we then have a modified de Broglie equation of motion%
\begin{equation}
\frac{dq}{d\eta}=v(q,\eta)_{\mathrm{reg}}=\frac{q\eta}{\eta^{2}+g^{2}}
\label{deBreg}%
\end{equation}
for the evolving degree of freedom $q=q(\eta)$.

For any function $g^{2}(\eta)$, our initial equilibrium Gaussian $\rho
(q,\eta_{i})=\left\vert \psi(q,\eta_{i})\right\vert ^{2}$ of width $\Delta
_{i}$ (and zero mean) evolves into a final nonequilibrium Gaussian of width%
\begin{equation}
D_{f}=X_{fi}\Delta_{i} \label{Df}%
\end{equation}
(and zero mean) where%
\begin{equation}
X_{fi}\equiv\exp\left(  \int_{\eta_{i}}^{\eta_{f}}d\eta\frac{\eta}{\eta
^{2}+g^{2}}\right)  \ . \label{Xfi}%
\end{equation}
To see this, consider a trajectory $q=q(\eta)$ that begins at $q_{i}%
=q(\eta_{i})$ and ends at $q_{f}=q(\eta_{f})$. A simple integration of
(\ref{deBreg}) yields%
\begin{equation}
q_{f}=q_{i}X_{fi}\ . \label{qf}%
\end{equation}
Furthermore, for the evolving distribution $\rho(q,\eta)$ we necessarily have
\[
\rho(q_{f},\eta_{f})dq_{f}=\rho(q_{i},\eta_{i})dq_{i}%
\]
(since trajectories beginning in a neighbourhood $dq_{i}$ of $q_{i}$ end in a
neighbourhood $dq_{f}$ of $q_{f}$). Because $\rho(q_{i},\eta_{i})=\left\vert
\psi(q_{i},\eta_{i})\right\vert ^{2}$ (by assumption), from (\ref{qf}) we then
have the final distribution%
\begin{equation}
\rho(q_{f},\eta_{f})=\frac{1}{X_{fi}}\left\vert \psi\left(  q_{f}/X_{fi}%
,\eta_{i}\right)  \right\vert ^{2}\ .
\end{equation}
This is indeed a Gaussian of width $D_{f}=X_{fi}\Delta_{i}$ and zero mean.

It is also straightforward to show that in this model we always obtain a final
super-quantum width%
\begin{equation}
D_{f}>\Delta_{f}\ .
\end{equation}
To see this, note that from (\ref{Delta}) the final equilibrium width
$\Delta_{f}$ may be written as%
\begin{equation}
\Delta_{f}=\Delta_{i}\sqrt{\frac{\eta_{f}^{2}+1/k^{2}}{\eta_{i}^{2}+1/k^{2}}%
}\ .\label{deltaf}%
\end{equation}
From (\ref{g2}) we have $g^{2}>1/k^{2}$ in the interval $(\eta_{i},\eta_{f})$
(assuming that $\alpha$ does not always vanish). From (\ref{Xfi}) we then have
(noting that $\eta<0$)%
\begin{equation}
X_{fi}>\exp\left(  \int_{\eta_{i}}^{\eta_{f}}d\eta\frac{\eta}{\eta^{2}%
+1/k^{2}}\right)  =\sqrt{\frac{\eta_{f}^{2}+1/k^{2}}{\eta_{i}^{2}+1/k^{2}}}%
\end{equation}
and so indeed $D_{f}>\Delta_{f}$. Thus, according to this model, the
time-dependent regulator always generates a power excess in the primordial
perturbations (in the relevant region of $k$-space).

From (\ref{Df}) and (\ref{deltaf}), at the final time $\eta_{f}$ we have a
nonequilibrium function%
\begin{equation}
\xi(k)\equiv\frac{D_{f}^{2}}{\Delta_{f}^{2}}=\left(  \frac{\eta_{i}%
^{2}+1/k^{2}}{\eta_{f}^{2}+1/k^{2}}\right)  X_{fi}^{2}\ ,\label{ksicalc}%
\end{equation}
with $X_{fi}$ given by (\ref{Xfi}). For any given regularisation -- specified
by $\alpha(\eta)$, or equivalently by $g^{2}(\eta)$ -- we may calculate
$X_{fi}$ and so find $\xi(k)$.

For example, let us consider a simple quadratic form%
\begin{equation}
g^{2}=(a-1)\eta^{2}+b\eta+c\label{g2quad}%
\end{equation}
with constants $a$, $b$ and $c$ chosen so that $g^{2}(\eta_{i})=g^{2}(\eta
_{f})=1/k^{2}$ (since $\alpha(\eta_{i})=\alpha(\eta_{f})=0$). For the case
$b^{2}<4ac$ we find%
\begin{equation}
\xi(k)=\left(  \frac{\Delta_{f}}{\Delta_{i}}\right)  ^{2(1-a)/a}\exp\left(
\gamma_{f}-\gamma_{i}\right)  \ ,\label{ksiquad}%
\end{equation}
where%
\begin{equation}
\gamma\equiv-\frac{2b}{a}\frac{1}{\sqrt{4ac-b^{2}}}\tan^{-1}\left(
\frac{2a\eta+b}{\sqrt{4ac-b^{2}}}\right)  \ .
\end{equation}
The constants $a$, $b$, $c$ may equally be written in terms of $\eta_{i}$,
$\eta_{f}$ and a constant $d$:%
\[
a=1-d\ ,\ b=(\eta_{i}+\eta_{f})d\ ,\ c=1/k^{2}-\eta_{i}\eta_{f}d\ .
\]
The dependence on $k$ is contained in $c$.

These illustrative examples serve as a starting point. One could of course
consider other choices for the regulator function and explore the extent to
which the results depend on the choice made.

\section{Trans-Planckian phenomenology and the CMB}

During the inflationary era an inflaton perturbation $\phi_{\mathbf{k}}$
generates a curvature perturbation $\mathcal{R}_{\mathbf{k}}\propto
\phi_{\mathbf{k}}$ (after the mode exits the Hubble radius), which in turn
generates the CMB angular power spectrum \cite{LL00}%
\begin{equation}
C_{l}=\frac{1}{2\pi^{2}}\int_{0}^{\infty}\frac{dk}{k}\ \mathcal{T}%
^{2}(k,l)\mathcal{P}_{\mathcal{R}}(k)\ , \label{Cl2}%
\end{equation}
where $\mathcal{T}(k,l)$ is the transfer function and%
\begin{equation}
\mathcal{P}_{\mathcal{R}}(k)\equiv\frac{4\pi k^{3}}{V}\left\langle \left\vert
\mathcal{R}_{\mathbf{k}}\right\vert ^{2}\right\rangle \label{PPS}%
\end{equation}
is the primordial power spectrum. From (\ref{ksi}) the nonequilibrium power
spectrum may be written as%
\begin{equation}
\mathcal{P}_{\mathcal{R}}(k)=\mathcal{P}_{\mathcal{R}}^{\mathrm{QT}}%
(k)\xi(k)\ , \label{xi2}%
\end{equation}
where $\mathcal{P}_{\mathcal{R}}^{\mathrm{QT}}(k)$ is the quantum-theoretical
or equilibrium power spectrum. Measurements of $C_{l}$ may be used to set
bounds on the deviation of $\xi(k)$ from $1$ \cite{AV10}.

Primordial quantum nonequilibrium is quantified by the function $\xi(k)$.
Extensive numerical studies of relaxation during a pre-inflationary era
indicate that $\xi(k)$ will take the form of an inverse-tangent -- with a
power deficit $\xi<1$ at small $k$ and with $\xi\simeq1$ at large $k$ -- where
the deficit is caused by incomplete relaxation at long wavelengths
\cite{CV13,CV14}. A large-scale power deficit has been found in data gathered
by the \textit{Planck} mission \cite{PlanckXV}. The magnitude and location of
the deficit are broadly consistent with pre-inflationary relaxation
suppression \cite{CV14}. But whether or not the predicted function $\xi(k)$ is
supported by the data remains to be seen \cite{PVV14}.

In this paper we are concerned with a different scenario. Instead of
considering relic nonequilibrium from earlier times, we are exploring the
possibility that nonequilibrium is created during inflation by novel effects
at the Planck scale. Nonequilibrium would then be expected to set in at
wavenumbers $k$ larger than some critical value $k_{\mathrm{c}}$ or at
wavelengths smaller than $\lambda_{\mathrm{c}}=2\pi/k_{\mathrm{c}}%
=\lambda_{\max}$ -- where modes of wavelength longer than $\lambda_{\max}$
were never sub-Planckian during the inflationary phase (see Figure 1).

For a region of $k$-space that is potentially subject to trans-Planckian
effects, we may attempt to predict features of the nonequilibrium function
$\xi(k)$. Our example of a time-dependent regulator leads to the form
(\ref{ksiquad}) for $\xi(k)$. This result depends on our simple choice --
defined by (\ref{finitedelta}) and (\ref{g2quad}) -- for the regulator. At
present we have no theoretical foundation for the regulator, which we have
introduced as an effective description of new and unknown physics at the
Planck scale. Therefore any test of the predicted modification of the power
spectrum by the function $\xi(k)$ may be seen as constraining the regulator
function. If one does adopt our simple choice of regulator, the parameters
appearing in the resulting expression (\ref{ksiquad}) for $\xi(k)$ are of
course unknown but in principle the general form of this function could be
supported (or not) by the data. This would require performing a best-fit to
the data, with the parameters in (\ref{ksiquad}) freely varying, to find out
if the fit is statistically significant or not. This is a matter for future work.

We have noted that our model with a time-dependent regulator can generate only
a power excess ($\xi>1$) and never a deficit ($\xi<1$). It may then seem that
this model could never account for the long-wavelength deficit reported by the
\textit{Planck} mission \cite{PlanckXV}. However, if there is a general power
excess below a critical wavelength $\lambda_{\mathrm{c}}$, then it could
happen that when the measured CMB power spectrum is normalised it will appear
\textit{as if} there were a power deficit above the same critical wavelength
$\lambda_{\mathrm{c}}$. Should $\lambda_{\mathrm{c}}$ be comparable to the
Hubble radius $H_{0}^{-1}$ today, our model could then predict an effective
deficit in the observed region.

To delineate the region of $k$-space that is potentially subject to
trans-Planckian effects -- specifically, to correction of the power spectrum
by the factor $\xi(k)$ -- we may consider the following simple estimates.

If inflation begins at a time $t_{\mathrm{begin}}$ and ends at a time
$t_{\mathrm{end}}$, then with an inflationary Hubble parameter $H$ the number
of e-folds will be $N=H(t_{\mathrm{end}}-t_{\mathrm{begin}})$. The relevant
range of $k$ -- where trans-Planckian effects can occur in the inflationary
spectrum -- is determined by maximum and minimum wavelengths $\lambda_{\max}$
and $\lambda_{\min}$, where modes with comoving wavelengths larger than
$\lambda_{\max}$ were never sub-Planckian during inflation while modes with
comoving wavelengths smaller than $\lambda_{\min}$ do not exit the Hubble
radius before inflation ends (see Figure 1). Thus%
\begin{equation}
a_{\mathrm{begin}}\lambda_{\max}\simeq l_{\mathrm{P}}\label{lmax}%
\end{equation}
and%
\begin{equation}
a_{\mathrm{end}}\lambda_{\min}\simeq H^{-1}\ .\label{lmin}%
\end{equation}
We then have relevant wave numbers $k$ in the range $(2\pi/\lambda_{\max}%
,2\pi/\lambda_{\min})$.%

\begin{figure}
[ptb]
\begin{center}
\includegraphics[width=0.7\textwidth]%
{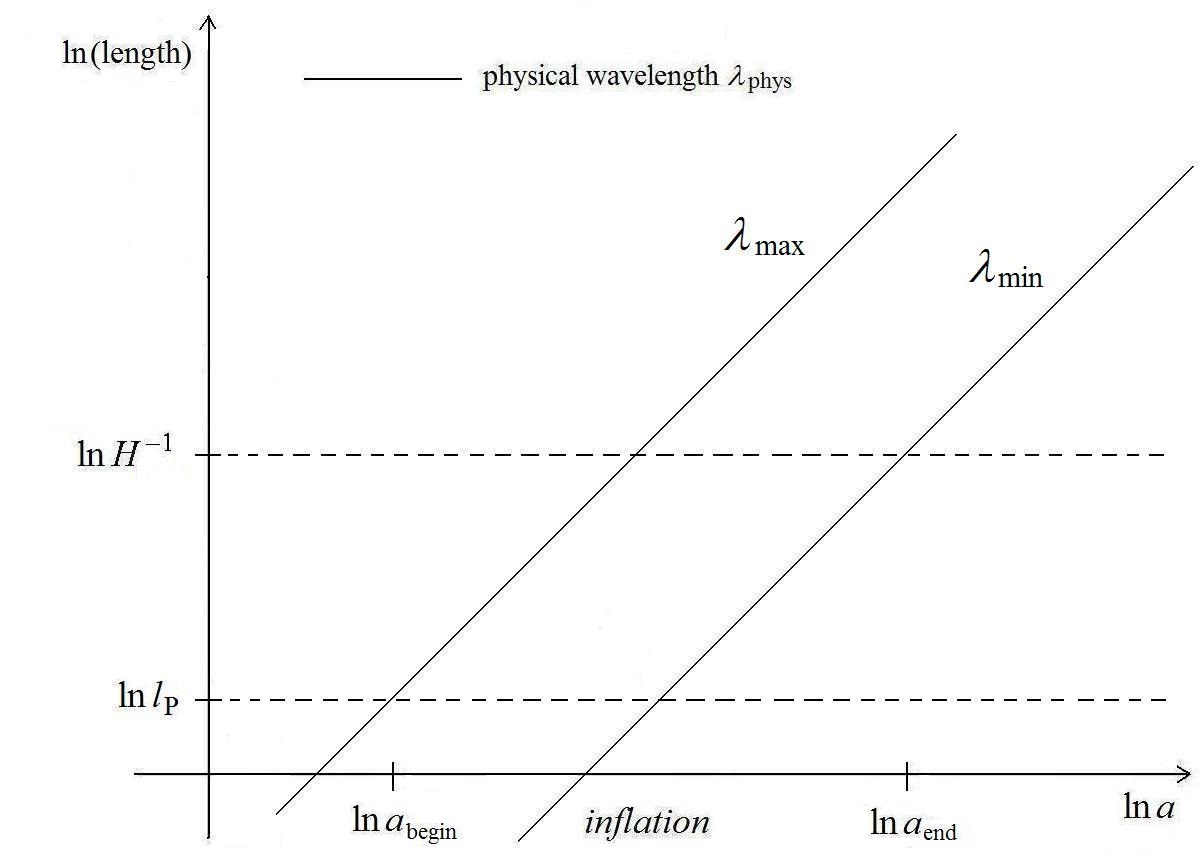}%
\caption{Physical wavelengths of trans-Planckian field modes during
inflation.}%
\end{center}
\end{figure}

In practice $\lambda_{\min}$ will be so small that we may as well take it to
be zero (see below). Thus the effects can set in at an ultraviolet cutoff
$\lambda_{\mathrm{c}}=\lambda_{\max}$ or%
\begin{equation}
\lambda_{\mathrm{c}}\simeq l_{\mathrm{P}}/a_{\mathrm{begin}}\ .
\label{cutoff1}%
\end{equation}
Modes with comoving wavelengths $\lambda>\lambda_{\mathrm{c}}$ were never
sub-Planckian during the inflationary phase and we can assume they are in
equilibrium (even if they were sub-Planckian during pre-inflation, we may
assume they relax to equilibrium during pre-inflation after they exit
$l_{\mathrm{P}}$).

We have $a_{\mathrm{begin}}=a_{\mathrm{end}}e^{-N}$ and so%
\[
\lambda_{\mathrm{c}}\simeq l_{\mathrm{P}}e^{N}/a_{\mathrm{end}}\ .
\]
If we neglect the expansion that takes place during the transition from
inflation to post-inflation, we can write $a_{\mathrm{end}}/a_{0}\simeq
T_{0}/T_{\mathrm{end}}$ (where $T_{0}$ is the temperature today and
$T_{\mathrm{end}}$ is the temperature at which inflation ends). Taking
$a_{0}=1$ it follows that%
\begin{equation}
\lambda_{\mathrm{c}}\simeq l_{\mathrm{P}}e^{N}(T_{\mathrm{end}}/T_{0}%
)\ .\label{cutoff2}%
\end{equation}
Using $l_{\mathrm{P}}\simeq10^{-33}\ \mathrm{cm}$ and writing $(1\ \mathrm{cm}%
)\simeq H_{0}^{-1}e^{-65}$ (where $H_{0}^{-1}\simeq10^{28}\ \mathrm{cm}$), we
then have%
\begin{equation}
\lambda_{\mathrm{c}}\simeq10^{-33}H_{0}^{-1}e^{(N-65)}(T_{\mathrm{end}}%
/T_{0})\ .\label{cutoff3}%
\end{equation}

Inflation can solve the horizon and flatness problems with a minimum number of
e-folds that is usually estimated to lie in the range $N_{\min}\simeq60-70$
\cite{PU09}. The actual number $N$ of e-folds could of course be much larger
than $N_{\min}$ \cite{Martin04}. The `reheating temperature' $T_{\mathrm{end}%
}$ depends on details of the reheating process such as the inflaton decay rate
\cite{PU09,Alla10}. Constraints from CMB data yield lower bounds on
$T_{\mathrm{end}}$ in the range $390\ \mathrm{GeV}-890\ \mathrm{TeV}$
(depending on the inflationary model) \cite{MR10}. Denoting the temperature at
the beginning of inflation by $T_{\mathrm{begin}}$, estimates for
$T_{\mathrm{end}}/T_{\mathrm{begin}}$ can range from $T_{\mathrm{end}%
}/T_{\mathrm{begin}}\sim1$ to $T_{\mathrm{end}}/T_{\mathrm{begin}}<<1$. We may
reasonably take $T_{\mathrm{begin}}$ to be of the same order of magnitude as
the energy scale $H\sim10^{16}\ \mathrm{GeV}\sim10^{-3}T_{\mathrm{P}}$
associated with the inflationary phase.

We may write (\ref{cutoff3}) as%
\[
\lambda_{\mathrm{c}}\simeq10^{-33}H_{0}^{-1}e^{(N-65)}(T_{\mathrm{end}%
}/1\ \mathrm{TeV})(1\ \mathrm{TeV}/T_{0})\ .
\]
Using $T_{0}\sim10^{-4}\ \mathrm{eV}$ we then have%
\begin{equation}
\lambda_{\mathrm{c}}\simeq10^{-17}H_{0}^{-1}e^{(N-65)}(T_{\mathrm{end}%
}/1\ \mathrm{TeV})\ .\label{cutoff4'}%
\end{equation}

As an illustrative example, the estimate (\ref{cutoff4'}) yields an order of
magnitude $\lambda_{\mathrm{c}}\sim H_{0}^{-1}$ if%
\begin{equation}
e^{(N-65)}(T_{\mathrm{end}}/1\ \mathrm{TeV})\sim10^{17}\ .
\end{equation}
This is consistent with the allowed parameter space. For example, we could
have $T_{\mathrm{end}}\sim10^{-3}T_{\mathrm{P}}$ or $T_{\mathrm{end}%
}/1\ \mathrm{TeV}\sim10^{13}$ together with $N\sim75$. To have much less than
the `maximal' reheating temperature requires a larger number of e-folds. For
example, to have $T_{\mathrm{end}}/1\ \mathrm{TeV}\sim1$ we would need
$N\sim105$.

As for $\lambda_{\min}$, from (\ref{lmin}) and using $a_{\mathrm{end}%
}=a_{\mathrm{begin}}e^{N}$ we may write%
\[
\lambda_{\min}\simeq(H^{-1}/l_{\mathrm{P}})e^{-N}(l_{\mathrm{P}}%
/a_{\mathrm{begin}})\simeq(H^{-1}/l_{\mathrm{P}})e^{-N}\lambda_{\max}\ .
\]
Thus $\lambda_{\min}$ is exponentially smaller than $\lambda_{\max}$. From
(\ref{cutoff2}) we have $\lambda_{\max}\simeq l_{\mathrm{P}}e^{N}%
(T_{\mathrm{end}}/T_{0})$ and so%
\begin{equation}
\lambda_{\min}\simeq H^{-1}(T_{\mathrm{end}}/T_{0})\ .
\end{equation}
For an inflationary energy scale $H\sim10^{16}\ \mathrm{GeV}$ we have
$H^{-1}\sim\hbar c/(10^{16}\ \mathrm{GeV})\simeq10^{-30}\ \mathrm{cm}$. As for
the ratio $T_{\mathrm{end}}/T_{0}$, taking a maximal value $T_{\mathrm{end}%
}\lesssim10^{16}\ \mathrm{GeV}$ (with $T_{\mathrm{end}}/T_{\mathrm{begin}%
}\lesssim1$) we have $\lambda_{\min}\lesssim10^{-1}\ \mathrm{cm}$. This is
indeed completely negligible and we may as well take $\lambda_{\min}=0$.

\section{Discussion and conclusion}

It is remarkable that trans-Planckian physics may be observable in the CMB,
enabling the above theoretical proposals to be constrained by experiment.
Trans-Planckian effects from more standard corrections to quantum field theory
(standard in the sense of remaining within the quantum formalism with the
usual Born rule) have been discussed by a number of authors. It appears that
oscillations in the primordial power spectrum are a generic prediction of such
models, though whether such features exist in the data remains a topic of
research \cite{BM13}. If one is willing to entertain the possibility of
trans-Planckian corrections to the Born rule, as suggested here, it will be
essential to find characteristic signatures that would enable such effects to
be distinguished from others. Our model with a time-dependent regulator yields
a power excess $\xi>1$. In our example $\xi$ was found to be given by the
expression (\ref{ksiquad}) as a function of $k$, though this result depends on
our simple choice of regulator for which we have as yet no theoretical
foundation. It may be hoped that deeper theoretical developments will lead to
firmer predictions, and that further analysis of the data (for example
best-fitting to power spectra corrected by factors of the form (\ref{ksiquad}%
)) will provide more detailed constraints on the kind of model proposed here.

Of the three arguments provided in Section 2 for quantum instability at the
Planck scale, we have focussed on a model with a time-dependent regularisation
of pilot-wave dynamics. Elsewhere \cite{PNAVinprep} we study
quantum-gravitational models for which there is no stable equilibrium state in
the deep quantum-gravity regime, as outlined in Section 2.2. It remains to be
seen what predictions could emerge from such models.

It was argued by Weiss \cite{Weiss85} that, in a quantum field theory with an
ultraviolet cutoff at a fixed physical lengthscale, on an expanding background
the number of field modes required to describe the physics will increase with
time. In effect, new degrees of freedom are born as the universe expands.
According to inflationary cosmology, these `new modes' could make an
observable contribution to the CMB spectrum \cite{BM13}. From a pilot-wave
perspective, the Born rule is a contingency and so there seems to be no
particular reason for why new modes should begin in a state of quantum
equilibrium. One might attempt to derive an estimate for the magnitude of the
nonequilibrium ratio $\xi$ for newly-born modes from an information-theoretic
argument. The `hidden-variable entropy' $S_{\mathrm{hv}}(k)$ of a field mode
$k$ may be taken to be minus the $H$-function (\ref{Hfn}) for the mode
\cite{AV10}. As a mode exits the Planck radius, it might be said that a new
degree of freedom is being created. Without attempting a proper justification,
we could assume that the hidden-variable entropy of these new degrees of
freedom will be given by $S_{\mathrm{hv}}\sim-\ln2$. One could think of this
(negative) entropy as being generated so as to `compensate' for the creation
of a new degree of freedom.\footnote{The relationship between $S_{\mathrm{hv}%
}$ and the usual von Neumann entropy is not fully understood; for a discussion
see refs. \cite{AV04,AVbook}.} For Gaussian packets $\rho$ and $\left\vert
\psi\right\vert ^{2}$ of respective widths $D$ and $\Delta$, we find
$S_{\mathrm{hv}}=-H=\frac{1}{2}\left(  1-\xi+\ln\xi\right)  $ where $\xi
=D^{2}/\Delta^{2}$. If we indeed assume that $S_{\mathrm{hv}}\sim-\ln2$, we
have (in the region of $k$-space where such effects could be relevant)
$1-\xi+\ln\xi\sim-2\ln2$. This yields two solutions, $\xi\sim0.1$ and $\xi
\sim3.7$, with respective sub-quantum and super-quantum widths. Something more
is needed to select one value over the other. Whether a rigorous argument can
be constructed along these lines remains to be seen.

In the context of pilot-wave theory, it is natural to ask why our universe
today is (at least to a good approximation) in a state of quantum equilibrium
while at the same time being in a state that is far from thermal equilibrium.
According to our current understanding, we observe thermal nonequilibrium
today because in the early universe gravitation amplified the small
inhomogeneities in temperature and energy density, leading to the formation of
large-scale structure \cite{Pad93}. Were it not for this peculiarity of
gravitation, our universe would now be in a state of global thermal
equilibrium. In contrast, we do observe global quantum equilibrium today: all
systems we have access to have been found to obey the Born rule (to high
accuracy). It might then seem that there can be no quantum analogue of the
gravitational amplification of thermal fluctuations. However, according to our
proposals, there can be circumstances -- albeit at the Planck scale -- in
which gravitation drives systems away from quantum equilibrium.\footnote{In
refs. \cite{AV04,AV07} it was suggested that such effects could exist at
macroscopic scales for entangled states straddling the event horizon of an
evaporating black hole.} If such effects do exist in the very early universe,
quantum nonequilibrium will be present at very early times -- even if the
universe began in a state of quantum equilibrium.

The proposals made in this paper arguably strengthen the parallels between
quantum and thermal fluctuations which have been noted by some authors in a
gravitational context. In particular, it has been argued that in the presence
of gravitation there is no invariant distinction between quantum and thermal
fluctuations \cite{Sciama81,Smolin86}. After several decades, possible deep
connections between quantum theory, gravitation and statistical physics remain tantalising.

If inflationary cosmology does indeed open an empirical window onto physics at
the Planck scale and beyond, then as well as considering the numerous
modifications of high-energy physics that have been proposed in the literature
we should take into account the possibility that quantum theory itself could
break down in such extreme conditions. It would therefore also be of interest
to explore inflationary collapse models \cite{PSS06,MVP12,LSS12,CPS13} in a
trans-Planckian context. We have advanced arguments suggesting that quantum
mechanics is unstable at the Planck scale, and we have provided an
illustrative model that makes use of the contingent status of the Born rule in
the de Broglie-Bohm formulation of quantum theory. We suggest that these
considerations are due for further development, and that the Born rule should
take its place alongside other basic features of modern physics that may be
questioned in the deep high-energy regime potentially probed by inflationary cosmology.

\textbf{Acknowledgements}. This research was funded jointly by the John
Templeton Foundation and Clemson University.

\end{document}